# CNN Classifier for Just-in-Time Woodpeckers Detection and Deterrent

Alexander Greysukh
agreysukh@gmail.com

## Abstract

Woodpeckers can cause significant damage to homes, especially in suburban areas. There are a number of preventing and repelling methods including passive decoys, though these may only provide temporary relief. Subsequently, it may be more efficient to implement a woodpecker deterrent, such as motion, light, sound, or ultrasound that would be triggered by detection of woodpecker signature drumming. To detect the typical 25 Hz drumming frequency, sampling periods under 10 milliseconds with frequent FFTs are required with considerable computational costs. An in-hardware spectrum analyzer may avoid these costs by trading off frequency for time resolutions. The trained model converted to TF Lite Micro, ported to an MCU, and identifies a variety of the prerecorded woodpecker drumming. The plan is to integrate the prototype with a deterrent device making it a completely autonomous solution.

## Keywords

## Introduction

Property damage caused by woodpeckers is a well-known problem causing substantial repair costs [1]. An autonomous drumming detection device combined with a deterring method might be an effective solution to the issue.

## Implementation

This approach is based on periodic analysis of sound spectrograms with a CNN classifier. The classifier was trained with a variety of drumming sounds and other interfering signals including ambient noises. Woodpecker drumming d lasts up to a couple of seconds with frequencies close to 25 Hz. Tor resolve these drumming patterns on the spectrograms, 5 milliseconds sampling within 3 seconds sliding window was sufficient. Running inference every 1 second allowed capturing of shorter drumming. To avoid computationally expensive FFT at each time point, a 7-band in-hardware spectra analysis was used. It made a tradeoff of frequency resolution for, more essential, time resolution.

The data sets were built by audio capturing pre-recorded drumming for several woodpecker species. All the preprocessing, including z-score normalization, was done on the board, and serialized spectrograms were sent to a custom annotation application. The spectrograms were collected in separate files along with the metadata. The TensorFlow and Keras libraries were used to generate the data sets and to train the serial CCN model. Fig 1 illustrates the processing workflow. Fig 2 shows typical drumming spectrograms. Table 1 summarizes the operational parameters.

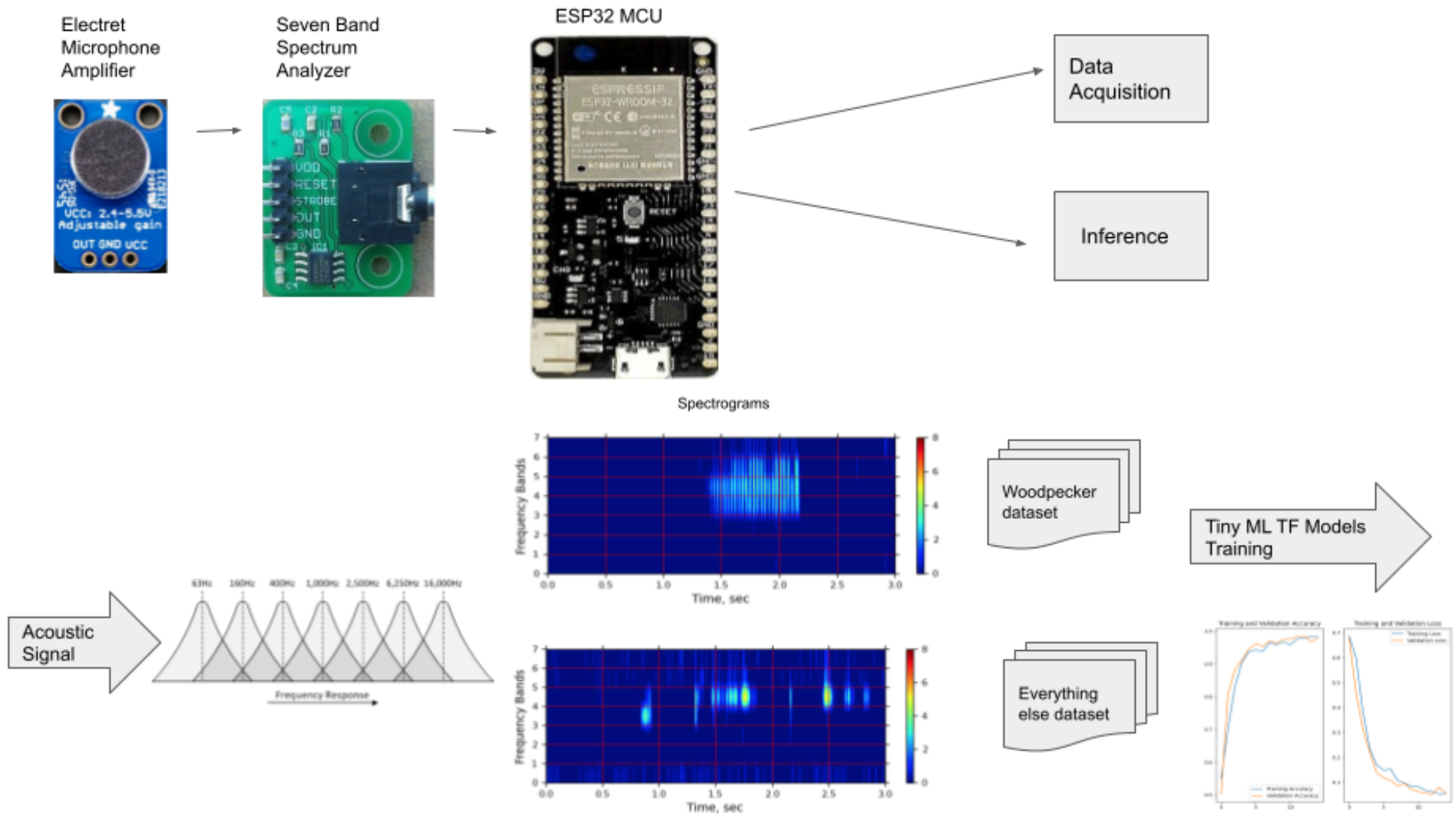

Fig 1. Hardware components and processing steps. The same board is used for data sets acquisition and for operation.

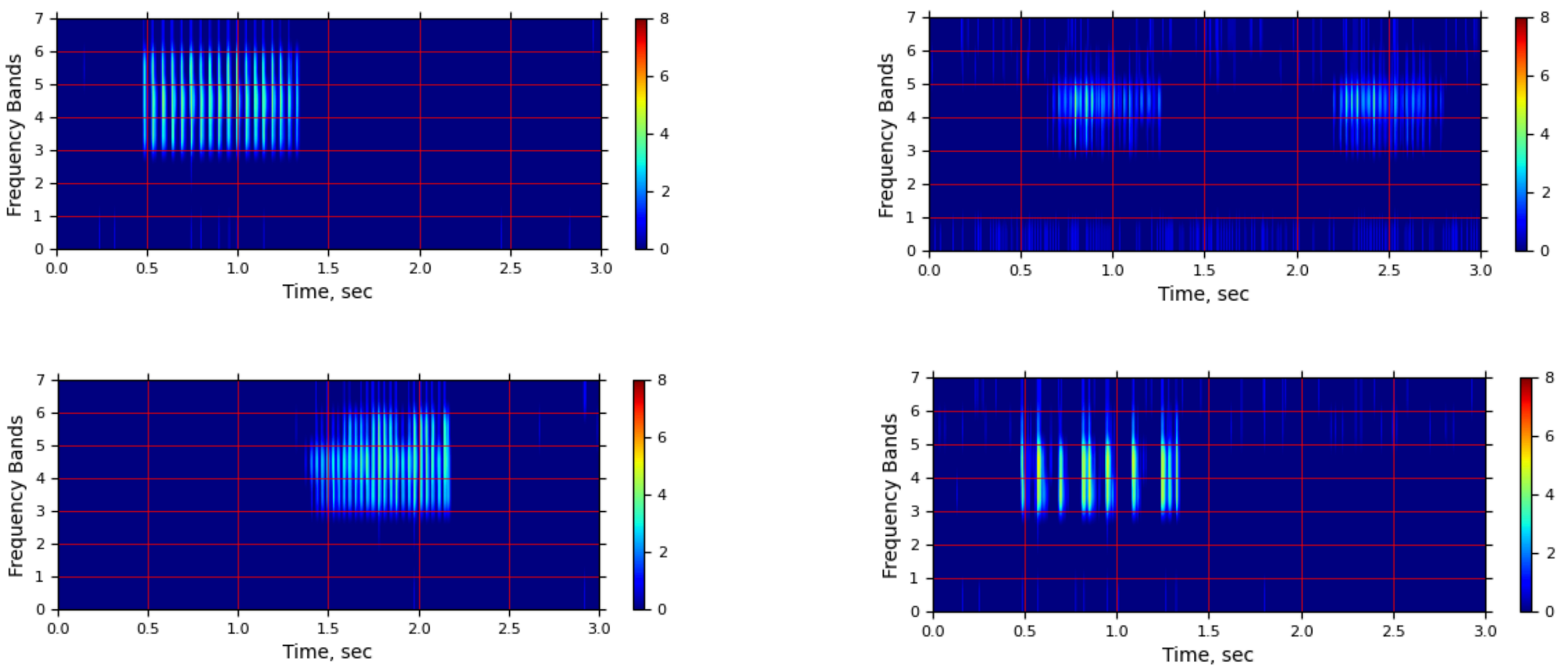

Fig 2.  Typical drumming spectrograms.

| Parameter | Value |
| --- | --- |
| **Sliding window** | 3 seconds |
| **Sampling period** | 5 milliseconds |
| **Inference period** | 1 second |
| **Spectrogram preprocessing duration** | ~5 milliseconds |
| **Model interpretation duration** | ~70 milliseconds |
| **Spectra analyzer bands** | 63Hz, 160Hz, 400Hz, 1kHz, 2.5kHz, 6.25kHz and 16kHz. |
| **Power consumption** | ~100 milliwatts |

Table 1. Operational parameters

# ML Model

The basic image classification CNN model was the first natural choice for the problem. It was adjusted for the spectrogram-specific shapes and trained with the ~750 files. Here is the model summary and an example of the training process:

```
Shape:  (8, 600, 7) (8, 1)
Model: "sequential"
_________________________________________________________________
Layer (type)                Output Shape              Param #
=================================================================
rescaling (Rescaling)       (None, 600, 7, 1)         0
_________________________________________________________________
conv2d (Conv2D)             (None, 600, 7, 4)         40
_________________________________________________________________
max_pooling2d (MaxPooling2D) (None, 200, 3, 4)        0
_________________________________________________________________
conv2d_1 (Conv2D)           (None, 200, 3, 8)         296
_________________________________________________________________
max_pooling2d_1 (MaxPooling2 (None, 67, 1, 8)         0
_________________________________________________________________
conv2d_2 (Conv2D)           (None, 67, 1, 16)         1168
_________________________________________________________________
max_pooling2d_2 (MaxPooling2 (None, 23, 1, 16)        0
_________________________________________________________________
dropout (Dropout)           (None, 23, 1, 16)         0
_________________________________________________________________
flatten (Flatten)           (None, 368)               0
_________________________________________________________________
dense (Dense)               (None, 32)                11808
_________________________________________________________________
dense_1 (Dense)             (None, 2)                 66
=================================================================
```

Total params: 13,378
Trainable params: 13,378
Non-trainable params: 0

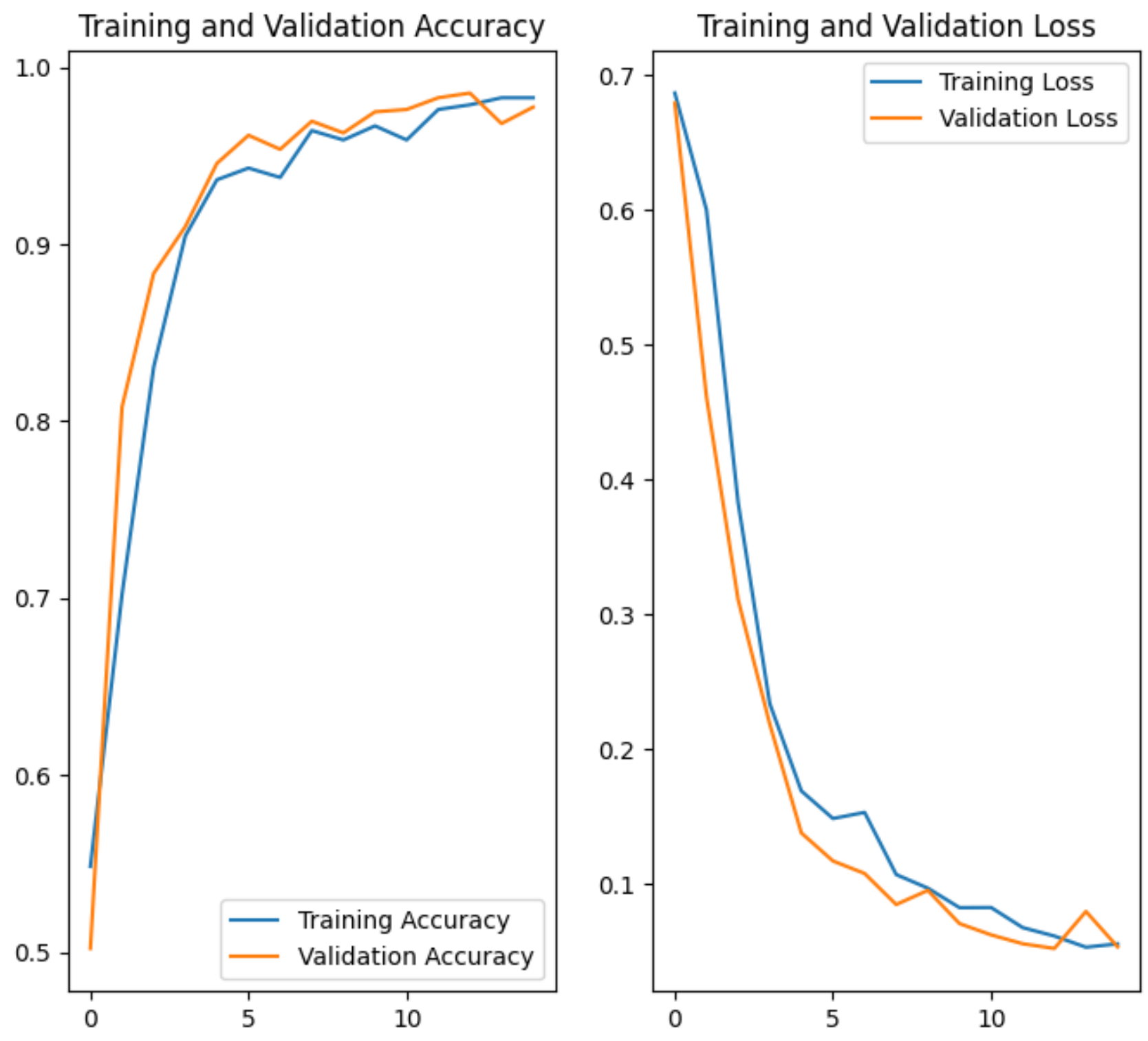

Fig 3. Training process for the model. Horizontal axis is epoch number.

Overall categorical accuracy ~98% reported after the training seems rather high. The model was converted to TFLite and to C++ without quantization as the first step.

# Schematics

An audio sensor, electret microphone amplifier with adjustable gain ( MAX4466 ) connected to a 7-band spectrum analyzer (MSGEQ7). The analyzer controlled from the MCU and the bands' amplitudes are passed to the ADC for each sampling period. ESP32 based board is used for the prototype, but it could be any other suitable MCU. An OLED display (SSD1306) shows operational status.

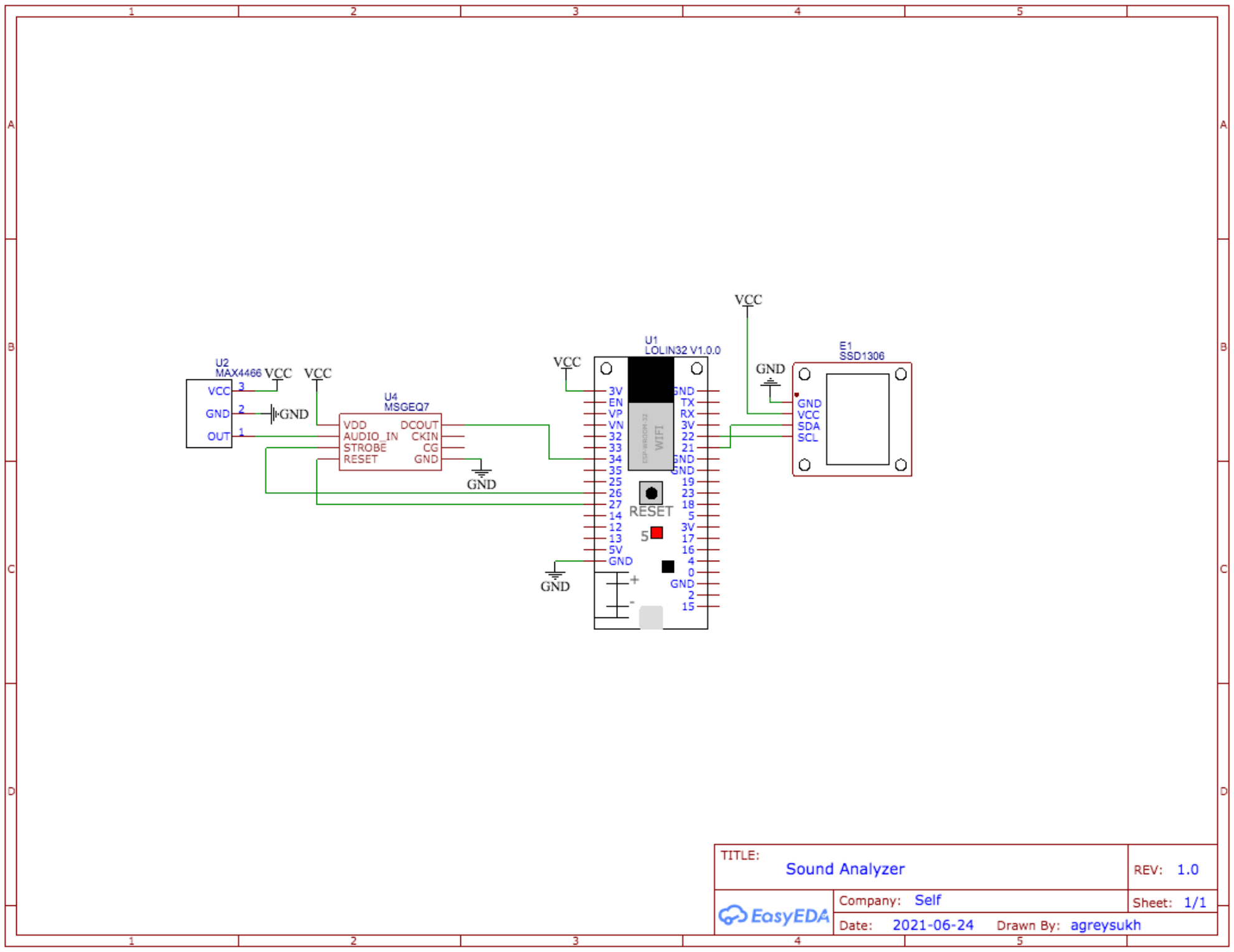

Fig 4. Schematics diagram.

# Summary

The prototype identifies pre-recorded woodpecker drumming with a low false positives rate. Several optimizations and enhancements may be required to make the classifier ready for testing in the field, including:

- A proximity sensor to save power (alternatively, one could wake up the classifier only when the ambient signal level is above a certain threshold).
- Solar battery charger.
- More sensitive directional microphone.

After integration with a repelling solution, this prototype could serve as a foundation for future development and productization.

# Hardware Components

Electret Microphone Amplifier - MAX4466 with Adjustable Gain
MSGEQ7 Seven Band Spectrum Analyzer Breakout Board Mono/Stereo (Mono)
LOLIN 32 MCU
OLED Monochrome Display SSD1306

## Software

TFLite and Keras C++ and Python libraries
PlatformIO IDE, Arduino, circular buffer, and device drivers libraries
PyCharm with IntelliJ IDE and Python 3.9 including multiple libraries

## Code

Will be published on GitHub